# Synthesis of mixed hypermetallic oxide BaOCa$^+$ from laser-cooled reagents in an atom-ion hybrid trap

**Authors:** Prateek Puri[1], Michael Mills[1], Christian Schneider[1], Ionel Simbotin[2], John A. Montgomery, Jr.[2], Robin Côté[2], Arthur G. Suits[3], and Eric R. Hudson[1]

**Affiliations:**

[1]Department of Physics & Astronomy, University of California, Los Angeles, California 90095, USA

[2] Department of Physics, University of Connecticut, Storrs, Connecticut 06269, USA

[3] Department of Chemistry, University of Missouri, Columbia, Missouri, 65211

*Correspondence to: eric.hudson@ucla.edu

**Abstract:** Hypermetallic alkaline earth (M) oxides of formula MOM have been studied under plasma conditions that preclude insight into their formation mechanism. We present here the application of emerging techniques in ultracold physics to the synthesis of a mixed hypermetallic oxide, BaOCa$^+$. These methods, augmented by high-level electronic structure calculations, permit detailed investigation of the bonding and structure, as well as the mechanism of its formation via the barrierless reaction of Ca ($^3$P$_J$) with BaOCH$_3$$^+$. Further investigations of the reaction kinetics as a function of collision energy over the range 0.005 K to 30 K and of individual Ca fine-structure levels compare favorably with calculations based on long-range capture theory.




**One Sentence Summary:** Emerging techniques of ultracold physics permit synthesis of the novel mixed hypermetallic oxide $BaOCa^+$ and yield deep insight into the formation mechanism and kinetics.

**Main Text:**

Molecules usually contain their constituent atoms in well-defined ratios predicted by classical theories of valence. Hypervalent species, however, are well known, and provide an opportunity to look more deeply into chemical bonding *(1)* and to anticipate and predict new chemical species and structures that may have exotic or useful properties *(2)*. An interesting class of hypervalent molecules is the hypermetallic alkaline earth (M) oxides of form MOM. Theory reveals the bonding in these linear molecules as donation of an electron from each metal atom to the central O atom, resulting in a system in which the central atom is closed shell, inhibiting coupling between the radical centers on the terminal metal atoms. As a result, the singlet-triplet splitting is very small and its prediction sensitive to the level of theory applied. The hypermetallic alkaline earth oxide BeOBe and its cation have recently been investigated by Heaven and coworkers using a range of spectroscopic tools, augmented by high-level electronic structure calculations *(3-4)*. For BeOBe, the singlet was found to be the ground state, just 243 $cm^{-1}$ below the triplet. Theoretical predictions of bonding and structure have also been reported for MgOMg *(5)*, CaOCa *(6)*, and SrOSr *(7)*.

Given these properties, MOM molecules and their cations provide an opportunity to benchmark quantum chemical calculations and explore bonding in molecules containing M atoms in the +1 oxidation state, which have recently been produced and are expected to



be useful for inorganic synthesis *(8)*. For mixed hypermetallic oxides MOM', dramatic effects on the electronic structure, single-triplet splitting, and excited state spectra may be expected to result from breaking the metal atom symmetry, leading to unusual inorganic diradicaloid systems *(9)*. In addition, for the mixed cations the asymmetric hole distribution impacts both the dipole moment and bonding. All of these properties could be tuned through choice of metal atoms for applications such as nonlinear optics, materials science, or as synthetic intermediates *(10)*. One challenge to such investigations is to develop a means to synthesize these molecules under controlled conditions and probe the pathways leading to their formation. Cations are a natural first target for such investigations as they can be manipulated and detected with great ease and sensitivity.

Emerging techniques in ultracold physics are now being adapted to the study of chemical systems, bringing new capabilities to probe reaction dynamics and mechanisms *(11-13)*. Recently, a study of the reaction of conformers of 3-aminophenol with laser-cooled $Ca^+$ ions revealed a fascinating dependence on conformational state *(14)*; quantum state resolved collisions between OH and NO *(15)*, as well between $N_2^+$ and Rb *(16)*, have been observed; and quantum effects were found to have a major impact on state-resolved KRb reactions *(17)*. Reactions involving polyatomic reagents are a compelling target for such studies as these techniques may be used to cool these species into a limited number of quantum states, as well as provide precise control over reaction conditions. Here we describe use of a magneto-optical atom trap coupled to an ion trap and time-of-flight mass spectrometer to synthesize a mixed hypermetallic alkaline earth oxide, $BaOCa^+$. These methods, augmented by high-level electronic structure calculations, permit detailed



investigation of the properties of this molecule as well as the mechanism of its formation via the barrierless reaction of Ca ($^3P_J$) with $BaOCH_3^+$.

**Experimental Apparatus**

Several unique aspects of the experimental apparatus were crucial to the feasibility of this study *(18)*. The hybrid ultracold atom-ion trap, dubbed the MOTion trap (Fig. 1A) *(19-20)*, consists of a radio-frequency linear quadrupole ion trap (LQT), co-located magneto-optical trap (MOT) of Ca atoms, and radially-coupled time-of-flight mass spectrometer (ToF) *(21, 22)*. Due to the spatial overlap of the atoms and ions, ultracold collisions between the two species and the chemical reactions and quantum phenomena they give rise to can readily be observed.

The collision energy of the trapped ions and atoms in this study, defined as $E/k_B$, where $E$ is the kinetic energy of the collision complex and $k_B$ is the Boltzmann constant, ranged from 0.005 K to 30 K, depending on the size of the ion crystal loaded into the LQT. A standard $Ba^+$ crystal was used as a sympathetic coolant for other trapped ions (Fig. 1B). In a typical experimental sequence, $Ba^+$ ions were initially loaded into the LQT through laser ablation of a $BaCl_2$ target. From this initial sample, a small number of $BaOH^+$ and $BaOCH_3^+$ ions were created by the reaction of $Ba^+$ with $CH_3OH$ *(23)* introduced into the vacuum chamber at a pressure of $\sim 10^{-10}$ Torr. The $BaOCH_3^+$ molecules were translationally cooled by the $Ba^+$ crystal. Recent studies *(19)* indicate that collisions with the ultracold Ca atoms of the MOT should also cool their ro-vibrational internal degrees of freedom; however, absent spectroscopy of $BaOCH_3^+$, we assume an internal temperature of < 300 K, bounded by the temperature of our vacuum chamber. Once a sufficient number



of BaOCH$_3^+$ molecules were produced, BaOH$^+$ ions could be removed from the LQT by resonantly exciting their motion at a mass-specific secular frequency; afterwards, the purified sample was immersed in a radius r ≈ 0.6 mm cloud of three million Ca atoms at 0.004(2) K (Fig. 1B). After a variable immersion time, $t_i$, the voltages of the LQT were adjusted *(24)* to eject the ions into the ToF *(18)*, yielding mass spectra (Fig. 1C).

The ToF spectra indicated the formation of a reaction product with mass-to-charge ratio *m/z* of 193(1) amu, which is consistent with that of BaOCa$^+$ (193.9 amu). We confirmed the assignment by introducing a photodissociating laser into the LQT and analyzing the dissociation fragments of the molecule *(See Supplementary Materials, Fig. S2)*. Depending on which dissociation pathway was resonant with the laser, fragments were detected with mass-to-charge ratios of either 40(1) amu or 153.7(3) amu, consistent with Ca$^+$ and BaO$^+$, respectively (Fig. 1D).

**Electronic Structure Calculations**

To aid in the interpretation of the experimental results, electronic structure calculations were performed for the Ca + BaOCH$_3^+$ → BaOCa$^+$ + CH$_3$ reaction *(24, 25)*. Optimized geometries for BaOCH$_3^+$ and BaOCa$^+$ and their fragments were obtained from density functional theory (DFT) using the triple-zeta correlation consistent basis sets (cc-pwCVTZ on calcium and barium and cc-pVTZ on hydrogen, carbon, and oxygen) and the B3LYP density functional. Coupled cluster theory including single and double excitations with perturbative triples, denoted CCSD(T), was used to estimate thermochemical energy differences *(18)*.



To check the validity of the DFT geometries for this problem, the CCSD(T) energies of the stationary points were recalculated at geometries obtained from second-order Møller-Plesset (MP2) theory, and the changes in thermochemical energy differences were less than 1 kcal/mol. DFT and MP2 offer different approaches to the electron correlation problem, but they predict geometries of generally comparable accuracy. Discrepancies between them would be an indication that a higher level of theory should be used, but their agreement here suggests such methods are not warranted. The electronic structure calculations were performed using the Gaussian 09 and Molpro 2012 program packages *(25, 26)*.

The calculated results predicted the Ca ($^1S_0$) + BaOCH$_3^+$ $\rightarrow$ BaOCa$^+$ + CH$_3$ reaction to be exothermic by 5.3 kcal/mol at the CCSD(T)/cc-pVTZ level of theory. Most of the exothermicity resulted from a loss of vibrational zero point energy between reactants and products. At the more expensive CCSD(T)/cc-pV5Z level of theory, the heat of reaction increased to 8.4 kcal/mol. In mixed hypermetallic oxides, the electronic degeneracy of the metal atom locations is removed and our calculations predicted electron localization on the Ca atom due to its higher ionization potential (Fig. 2A). This conclusion was supported by a natural bond order analysis assigning partial charges of +1.67 to barium and +0.91 to calcium in BaOCa. Calculations also indicated the ion has a significantly larger permanent dipole moment (2.80 D) than neutral BaOCa (1.32 D), again supporting principal removal of Ba-centered electron density upon ionization. The first strong electronic transition in BaOCa$^+$ corresponds to transfer of this electron density from Ca to Ba. As this electron does not strongly participate in the molecular bonding, the associated Franck-Condon factors are moderately diagonal and may allow optical cycling and detection *(27)*. We



calculated the ionization energy of BaOCa to be 4.18 eV, slightly higher than in BaOBa (experimentally reported as 3.87 eV *(28)*) but closer to BaOBa than CaOCa, calculated to be 4.90 eV.

Calculations for neutral BaOCa predicted that, like BeOBe, it is a diradicaloid system with a similarly small singlet-triplet splitting of only 407 cm$^{-1}$ but with very different energies for the radical centers. The small singlet-triplet splitting in *neutral MOM' molecules* is a manifestation of the spin uncoupling on the metal centers. The reaction experimentally studied in this work produces the BaOCa$^+$ cation and a CH$_3$ coproduct, two doublets whose spins are uncorrelated, and thus, the singlet-triplet splitting vanishes and the potential energy surfaces are degenerate.

A calculation of the intrinsic reaction coordinate (IRC) leading from the transition state to reactants and products was performed (B3LYP/cc-pVTZ) along the Ca ground-state singlet surface (Fig. 2C) and revealed the existence of two bound BaOCH$_3$Ca$^+$ complexes, one in the entrance channel and one in the exit channel. These structures and their relative energies were further investigated at the more sophisticated CCSD(T)/cc-pVTZ level of theory (Fig. 2A), indicating the existence of a 10.2 kcal/mol barrier to the reaction *(18)*.

Finally, multi-configurational self-consistent field calculations were performed on all stationary points presented in Fig. 2A and verified that multi-reference effects do not play a significant role in the system, except possibly in the singlet transition state *(18)*. To this end, natural orbital analysis and a coupled cluster theory calculation, utilizing the singlet wavefunction with the two most significant configurations included, was performed and indicated a 3.4 kcal/mol increase in the singlet barrier height. This barrier still



precludes reaction along the singlet surface, and the calculation further verifies that multi-reference effects would not significantly alter the conclusions of our computational study.

**Experimental Search for Reaction Pathway**

Given that the predicted barrier is insurmountable at experimentally realized collision energies and that the tunneling probability through the barrier is negligible, we hypothesized the observed synthesis occurred through an electronically excited state of the Ca reactant. To test this explanation, we varied the Ca electronic state populations via control of the Ca MOT lasers *(18)* and measured the resultant changes in BaOCa$^+$ production. The excited state populations of the Ca atoms were determined from a rate equation model spanning 75 electronic states that incorporated the intensities and detunings of all near-resonant laser fields present in the MOT trapping volume *(29)*. The chemical reaction rate for the Ca + BaOCH$_3^+$ $\rightarrow$ BaOCa$^+$ + CH$_3$ reaction is given by $\Gamma = n_a k_t$, where $n_a$ is the Ca atom number density and $k_t$ the total reaction rate constant, which is found as $k_t = \sum_i p_i k_i$, where $p_i$ and $k_i$ are the population and reaction rate constant of the $i^{\text{th}}$ electronic state, respectively. The total reaction rate constant was experimentally measured by monitoring the amount of both BaOCH$_3^+$ and BaOCa$^+$ present in the LQT as a function of interaction time with a Ca MOT of known density. The solution of a differential equation incorporating all measured loss and production rates for each molecular ion due to photodissociation, chemical reactions, and background loss was then fit to the reaction kinetics data in order to determine $k_t$ *(18)*.

The experimentally measured reaction rate exhibited no statistically significant dependence on the population of the singlet Ca electronic states involved in the laser



cooling process, i.e. the *4s²* ¹S₀, *4s4p* ¹P₁, *4s5p* ¹P₁ and *3d4s* ¹D₂ states *(18)*. This observation is consistent with preliminary theoretical calculations, which suggested that a reaction barrier, similar to that of the Ca (¹S₀) + BaOCH₃⁺ channel, exists on all of these singlet channels.

Studies have shown *(29, 30)* spin-forbidden optical transitions lead to the production of a small number of Ca atoms in the *4s4p* ³P_J states (Fig. 3C) in Ca MOTs. Though atoms in these metastable states are not trapped by the MOT force, they are continually produced, leading to a steady-state population in the trapping volume. Further, controlling the MOT lasers can vary the electronic populations in these states and reveal how they affect the reaction rate in a manner similar to studies of the singlet state. The observed reaction rate as a function of total population in the *4s4p* ³P_J states is shown in Fig. 2D, with a characteristic kinetics data set and the corresponding fitted solutions shown in the inset. Here, the linear dependence of the reaction rate constant on the *4s4p* ³P_J population was shown to be consistent with zero vertical intercept, suggesting that the observed formation of BaOCa⁺ initiates predominantly along the triplet Ca (³P_J) + BaOCH₃⁺ surface.

While non-adiabatic interactions from the excited singlet surfaces coupling to other electronic states could permit reaction despite the calculated barriers, the experimental observations indicate that these effects, if present, do not play a significant role *(18)*. Additionally, since the collected data is sensitive to reaction entrance channel, but not necessarily to the surface along which the reaction completed, events where coupling from the triplet surface to the singlet surface occurred and resulted into reaction would not be



experimentally distinguishable from reactions evolving exclusively along the triplet surface.

**Experimental Verification of Triplet Reaction Pathway**

In order to verify the Ca $^3P_J$ pathway of the reaction, we performed two additional experiments. First we measured the reaction rate of Ca atoms in a single internal quantum state, the $\left|{}^3P_2, m_J = 2\right\rangle$ state, by loading MOT atoms in this state into a magnetic trap and overlapping them with the ions. This experiment showed unequivocally that the reaction occurs between a Ca atom in the $^3P_J$ state and a $BaOCH_3^+$ ion. In the second experiment, we employed additional optical pumping lasers to populate only a single $^3P_J$ state during Ca MOT operation, enabling the extraction of fine-structure–resolved reaction rate constants for the $^3P$ states.

Under normal MOT operation, multiple energy levels in the laser cooling cycle are populated simultaneously. Although the triplet population data in Fig. 2D suggests the $^3P_J$ pathway of the reaction, it is possible that other electronic states may be contributing to the observed reaction through non-adiabatic processes. The magnetic trap, introduced above, provides a means to isolate a sample of triplet Ca atoms and ensure that reaction initiates on the triplet surface. The magnetic trap, a separate atom trap from the MOT whose non-optical trapping force is produced by the MOT field gradients, serves as a nearly pure reservoir of Ca triplet atoms since only atoms in the $\left|{}^3P_2, m_J = 2\right\rangle$ state have large enough magnetic moment to produce significant atomic trap densities *(18)*.



In the magnetic trapping experiment, ions were first initialized as described earlier. To ensure that reaction only occurred between the magnetically trapped Ca atoms and BaOCH$_3^+$ molecules, the voltages of the LQT were adjusted such that BaOCH$_3^+$ ions were first displaced from the center position of the MOT by ~3 mm, corresponding to a displacement of ~5 MOT radii, precluding background reactions from direct MOT-BaOCH$_3^+$ overlap. After the magnetic trap was loaded to capacity, the MOT was depleted by extinguishing the atom cooling beams, removing any background Ca MOT atoms from the magnetic trap region within ~ 5 ms. The endcap voltages were then adjusted to shuttle the ion crystal to the center of the magnetic trap, allowing it to react directly with a nearly pure sample of $|^3P_2, m_J = 2\rangle$ atoms for the duration of the magnetic trap lifetime (~500 ms), and this process was repeated up to 100 times for each ion crystal. Here, $m_J$ is defined with respect to the trap magnetic field direction, whereas the relative velocity vector defining the reaction is isotropically distributed, meaning the Ca $m_J$ sublevel is not controlled along the reaction quantization axis.

BaOCa$^+$ accumulation in the LQT was observed to increase with BaOCH$_3^+$ magnetic trap immersion time, whereas the chemical reaction rate for a control case, where an optical pumping laser was used to depopulate $^3P_2$ atoms throughout the experiment and thus deplete the magnetic trap, was consistent with zero (Fig. 3B). When $|^3P_2, m_J = 2\rangle$ atoms were present in the magnetic trap, a reaction rate constant of ~$10^{-9}$ cm$^3$/s was measured, consistent with the reaction rate measurement described earlier. Here, fluorescence imaging and spatial estimates of the magnetic trap derived from the magnetic field gradients of the MOT were used to estimate the $^3P_2$ atom number density needed for

the rate constant calculation. The uncertainty of this estimate prevents a more precise measurement of the reaction rate constant *(18)*.

Therefore, to find more accurate reaction rate constants and to resolve the rate constant for each of the $^3P_J$ fine-structure states, optical pumping lasers were used to deplete population from two $^3P_J$ levels simultaneously, isolating population in a single triplet state while reaction kinetics data were measured. All three measured fine-structure resolved reaction rate constants (Fig. 4D) were of order $10^{-9}$ cm$^3$/s, with the $^3P_1$ state exhibiting the largest rate constant value of 5.4(9) x $10^{-9}$ cm$^3$/s. These results are in reasonable agreement with predictions from a long range capture theory (see below).

**Triplet Surface Electronic Structure Calculations and Long-Range Capture Model**

Having concluded experimentally that the synthesis of BaOCa$^+$ occurs via the triplet channel, electronic structure calculations, as described earlier, were performed to characterize the Ca ($^3P$) + BaOCH$_3^+$ $\rightarrow$ BaOCa$^+$ + CH$_3$ reaction. Although the general features of the triplet potential energy surface leading to the two product doublet molecules were similar to those discussed above for the ground state, the transition state for reaction on the triplet surface (Fig. 2C) was calculated to be 25.4 kcal/mol (CCSD(T)/cc-pVTZ ) below the energy of the reactants (Fig. 2, A and B), meaning the reaction proceeds without barrier for each $^3P_J$ fine-structure state. The ground state and triplet potential surfaces both have entrance channel complexes that feature a strongly bent Ba-O-Ca backbone with the methyl attached to the oxygen while retaining the pyramidal *sp$^3$* configuration. The ground state exit channel shows a strongly bound complex with the methyl chemically bound to Ca, whereas the triplet exit channel minimum may be characterized as a van der Waals



type interaction between a planar methyl radical and the incipient BaOCa$^+$ molecule. This reaction shows very different dynamics on the singlet and triplet surfaces, but in contrast to the commonly seen case of a singlet atom inserting into a covalent bond *(31)*, here the triplet is more reactive as the singlet-triplet splitting is significant in the calcium atom but small at the transition state and in the product.

The predicted absence of a barrier suggested that the observed reaction rate could be estimated from long-range capture theory. To this end, we evaluated molecular potential curves (Fig. 4A) for Ca ($^3P_J$) in states (J, $m_J$) by considering both the long range R$^{-3}$ interaction associated with the quadrupole moment of the $^3P_J$ states and the usual R$^{-4}$ polarization potential *(32-34)*. The curves for $\pm m_J$ are identical, resulting in three distinct curves for J=2, two for J=1, and a single curve for J=0. The effect of the quadrupole moment is non-trivial, leading to barriers that reduce reaction rates for some channels or more attractive curves that increase the reaction rates for others.

To compute theoretical energy-dependent reaction rates, we employed a simple Langevin capture model *(35, 36)*. Fig. 4B shows the results for each (J, $|m_J|$) state as a function of the collision energy. An energy-dependent rate constant for each fine structure component J (Fig. 4C) was calculated by summing over the $m_J$ components. Whereas at collision energies greater than ~10 K the rate constants decrease with J, this trend shifts drastically at lower temperatures and even reverses for collision energies below ~1 K. To compare with experimental rate coefficients, these rates were averaged over the velocity distribution of the ions in the LQT. The results, which demonstrate reasonable agreement with experiment, are shown in Fig. 4D, where an uncertainty band based on a 10% range



in published theoretical values for the quadrupole moments and polarizabilities used in the molecular potential calculations is also included *(18, 37-39)*.

Finally, to directly probe for the existence of a barrier on the triplet surface, we monitored the collision energy evolution of the reaction rate constant. Since the micromotion energy in an ion trap scales with the spatial radial width of the ion crystal, average collision energies can be controlled by simply changing the size of ion crystals initially loaded into the LQT *(18)*. Using this method, we probed reaction rates at average collision energies ranging from 0.1 K to 30 K and compared the results (Fig. 4E) to the capture theory prediction weighted by the spatially-dependent energy distribution of the ions. As seen here, the measured reaction rate constant does not have a strong collision energy dependence over these temperatures and agrees with the capture theory calculation, indicating a barrierless reaction.

Further, in experiments with linear ion chains, $BaOCa^+$ formation was still observed at the lowest collision energies reached of ~0.005 K, confirming the absence of potential barriers to the reaction at temperatures near the ultracold regime. However, at these temperatures, the ion crystals used in the LQT are extremely small, and due to the large accumulation time needed for $BaOCH_3^+$ buildup and ToF measurement shot noise, accurate reaction rate data were experimentally inaccessible. Consequently, such temperatures were excluded from the kinetics data shown in Fig. 4E.



**Outlook**

Through precise control of entrance channels and fine-tuning of reaction energetics, from high temperature to the ultracold regime, techniques used here and elsewhere *(14-17)* offer promising platforms for extending the tools of ultracold physics to the study of high precision quantum chemical dynamics. Therefore, they are expected to enable a next generation of chemical studies in the quantum regime, providing opportunities to look more deeply into chemical bonding and to anticipate and predict new chemical species and structures that may have exotic or useful properties.



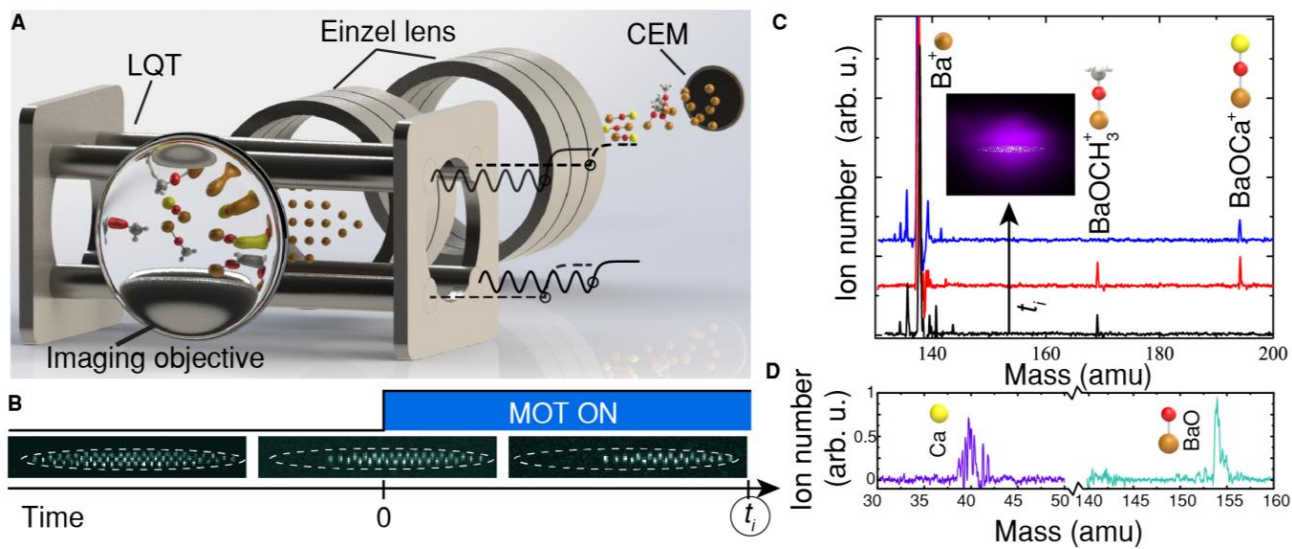

**Figure 1. Experimental schematic of the hybrid system and ToF apparatus.**

**(A)** A schematic of the experimental apparatus, including the LQT, the high voltage pulsing scheme (shown as solid and dashed lines), and the ToF. **(B)** An illustrative experimental time sequence that depicts initialization of a $Ba^+$ crystal, production of $BaOCH_3^+$ (visualized as dark ions in the crystal) through reactions with methanol vapor, and subsequent MOT immersion. **(C)** Sample mass spectra obtained after ejecting the LQT species into the ToF after various MOT immersion times, $t_i$, along with an inset depicting a superimposed fluorescence image of an ion crystal immersed in the Ca MOT. **(D)** Mass spectra of photofragmentation products collected after inducing photodissociation of $BaOCa^+$. The identified photofragments were used to verify the elemental composition of the product.



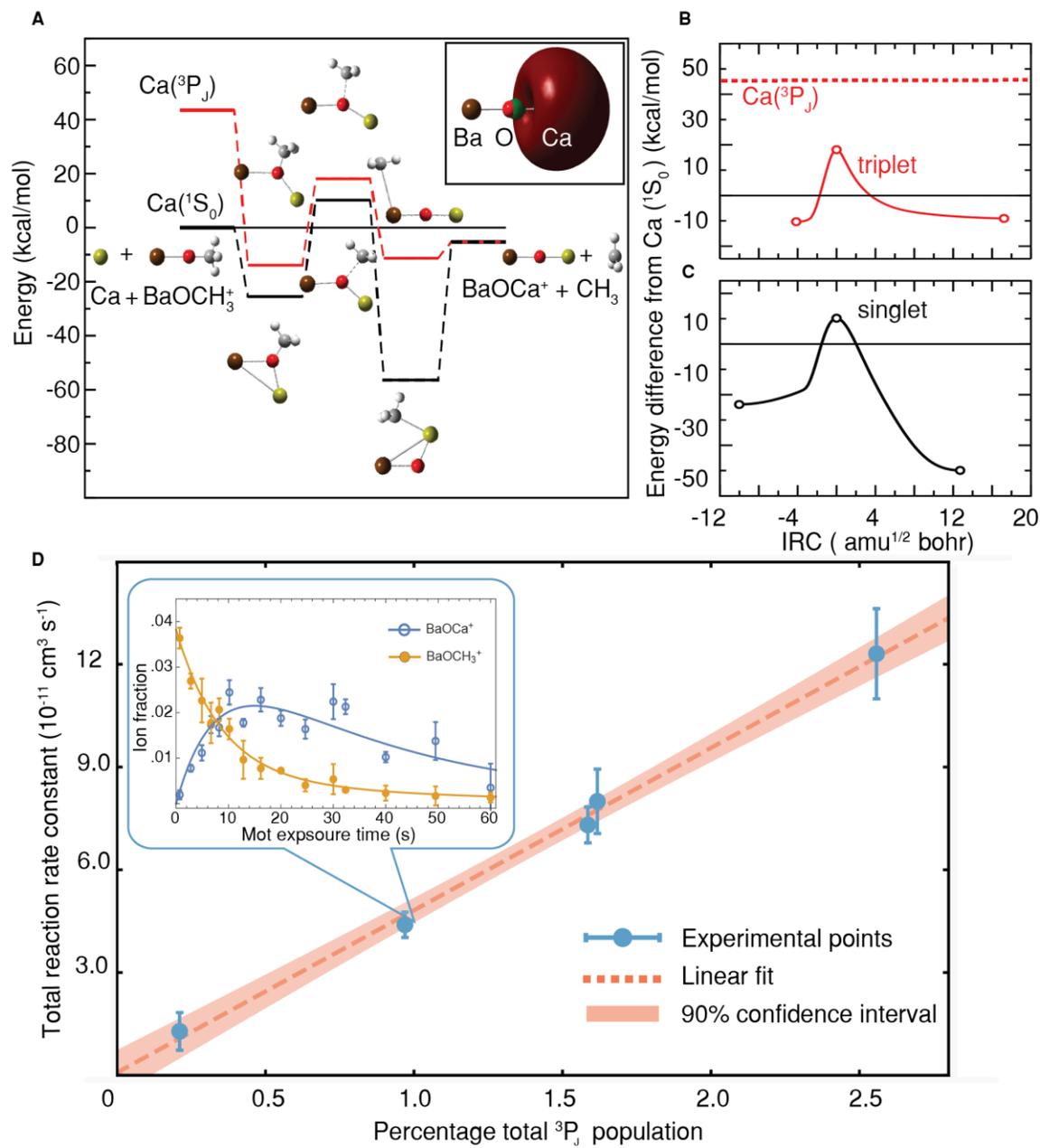

**A**

Ca($^3P_J$)

Ca($^1S_0$)

Ca + BaOCH$_3^+$

BaOCa$^+$ + CH$_3$

Energy (kcal/mol)

Ba  O  Ca

**B**

Ca($^3P_J$)

triplet

Energy difference from Ca ($^1S_0$) (kcal/mol)

**C**

singlet

IRC ( amu$^{1/2}$ bohr)

**D**

Total reaction rate constant ($10^{-11}$ cm$^3$ s$^{-1}$)

Ion fraction

BaOCa$^+$
BaOCH$_3^+$

Mot exposure time (s)

- ● - Experimental points
- ···· Linear fit
- ▬ 90% confidence interval

Percentage total $^3P_J$ population



**Figure 2. BaOCa⁺ production mechanism.**

**(A)** Energy of stationary points along the Ca $^1S_0$ (black) and $^3P_J$ (red) reaction pathways calculated at the CCSD(T)/cc-pV5Z level of theory. The corresponding energies for the singlet (triplet) pathway in kcal/mol are, from left to right, 0 (43.5), -25.5 (-13.9), 10.2 (18.1), -56.4 (-11.3), and -5.3 (-5.3). The presence of a barrier in the Ca $^1S_0$ pathway precludes reaction at low temperature, while the transition state in the triplet pathway is well below the energy of the reactants and does not prevent the exothermic reaction to BaOCa⁺ and $CH_3$. The geometries of the complexes at each stationary point are shown below (above) the singlet (triplet) pathway. The inset displays the linear geometry of the BaOCa⁺ molecule and its open shell highest occupied molecular orbital. **(B-C)** Energy along the IRC for both the singlet (B) and triplet (C) surfaces calculated at the B3LYP/cc-pVTZ level of theory. The circles correspond to the stationary points in (A), and all energies are given with respect to the ground state reactants. **(D)** Experimental total reaction rates plotted as a function of aggregate triplet Ca population, presented alongside a linear fit to the data (weighted by the reciprocal of the standard error squared) and its corresponding 90% confidence interval band. Experimental uncertainties are expressed at the one-sigma level. The inset shows the temporal evolution of both BaOCH$_3^+$ and BaOCa⁺ amounts, normalized by initial Ba⁺ number, in the LQT as a function of MOT exposure time as well as the solutions of differential equations globally fit to ~250 kinetic data points in order to extract reaction rate constants, with a reduced chi-square statistic of 1.03 specifying the goodness-of-fit to the displayed data set *(17)*.



**A**

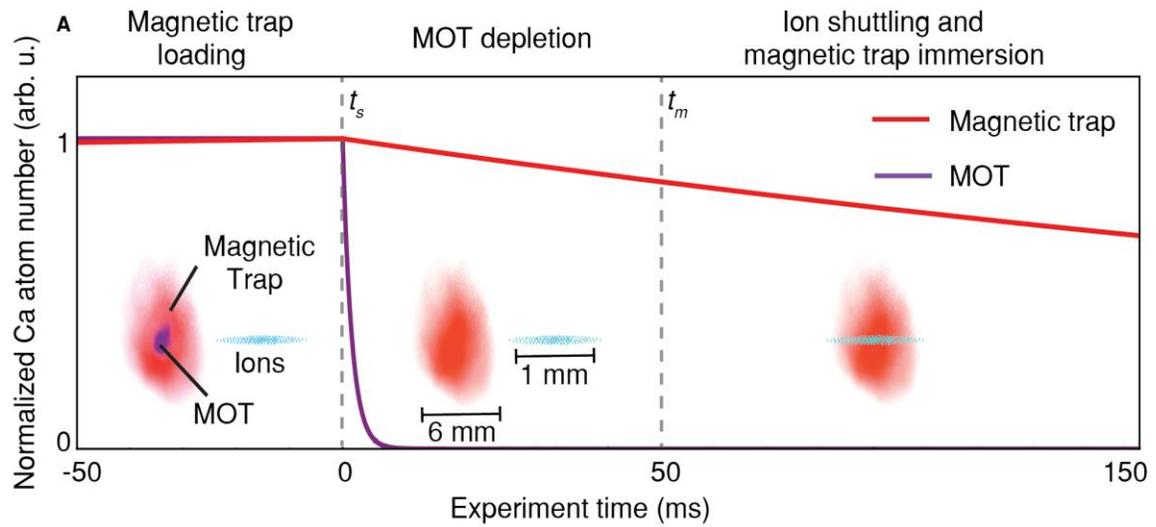

Magnetic trap loading — MOT depletion — Ion shuttling and magnetic trap immersion

$t_s$    $t_m$

Magnetic trap
MOT

Magnetic Trap
Ions
MOT

6 mm    1 mm

Normalized Ca atom number (arb. u.)

Experiment time (ms)

**B**

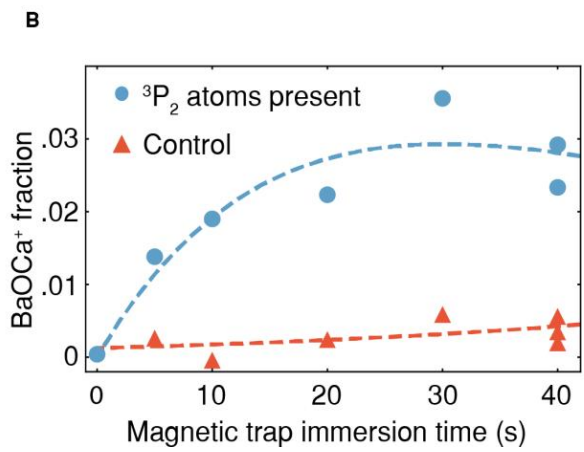

$^3P_2$ atoms present
Control

BaOCa$^+$ fraction

Magnetic trap immersion time (s)

**C**

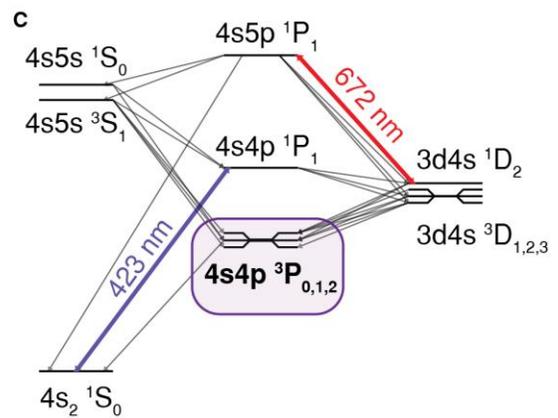

4s5s $^1S_0$
4s5p $^1P_1$
4s5s $^3S_1$
4s4p $^1P_1$
672 nm
3d4s $^1D_2$
3d4s $^3D_{1,2,3}$
4s4p $^3P_{0,1,2}$
423 nm
4s$_2$ $^1S_0$



**Figure 3. Production of BaOCa$^+$ through reaction with metastable magnetically trapped calcium.**

**(A)** The number of atoms (normalized by the initial atom amount in each trap) in both the magnetic trap and the MOT probed as a function of experiment time by monitoring the amount of fluorescence produced from each when illuminated with a near-resonant laser. A typical experimental time sequence is also presented, along with scaled false-color fluorescence images of both the atoms and ions for illustration. Approximate spatial scales, provided separately for the atom and ion images, are also displayed for reference. Ions are initially displaced from the MOT as the magnetic trap is loaded. At $t_s$, the atom cooling beams are extinguished to deplete MOT atoms from the magnetic trap region, and the LQT endcaps are subsequently adjusted at $t_m$ to overlap the ions with the center of the magnetic trap for roughly 500 ms, enabling BaOCH$_3^+$ reactions with Ca ($^3$P$_2$) atoms. **(B)** BaOCa$^+$ accumulation, expressed as a fraction of initial Ba$^+$ amount, plotted as a function of interaction time with the magnetic trap. A control case where a laser is used to depopulate the $^3$P$_2$ Ca level during magnetic trap loading is also presented. Fitted solutions to differential equations, obtained in the same manner as those in Fig. 3C, are presented alongside the data, and, after estimating the magnetic trap density, they yield reaction rate constants of 8(3) x $10^{-9}$ cm$^3$/s and 0(3) x $10^{-9}$ cm$^3$/s for the experimental case and the control, respectively *(17)*. **(C)** A level scheme for Ca including the relevant electronic states involved in the laser cooling process, with the reactive $^3$P$_{0,1,2,}$ states highlighted.



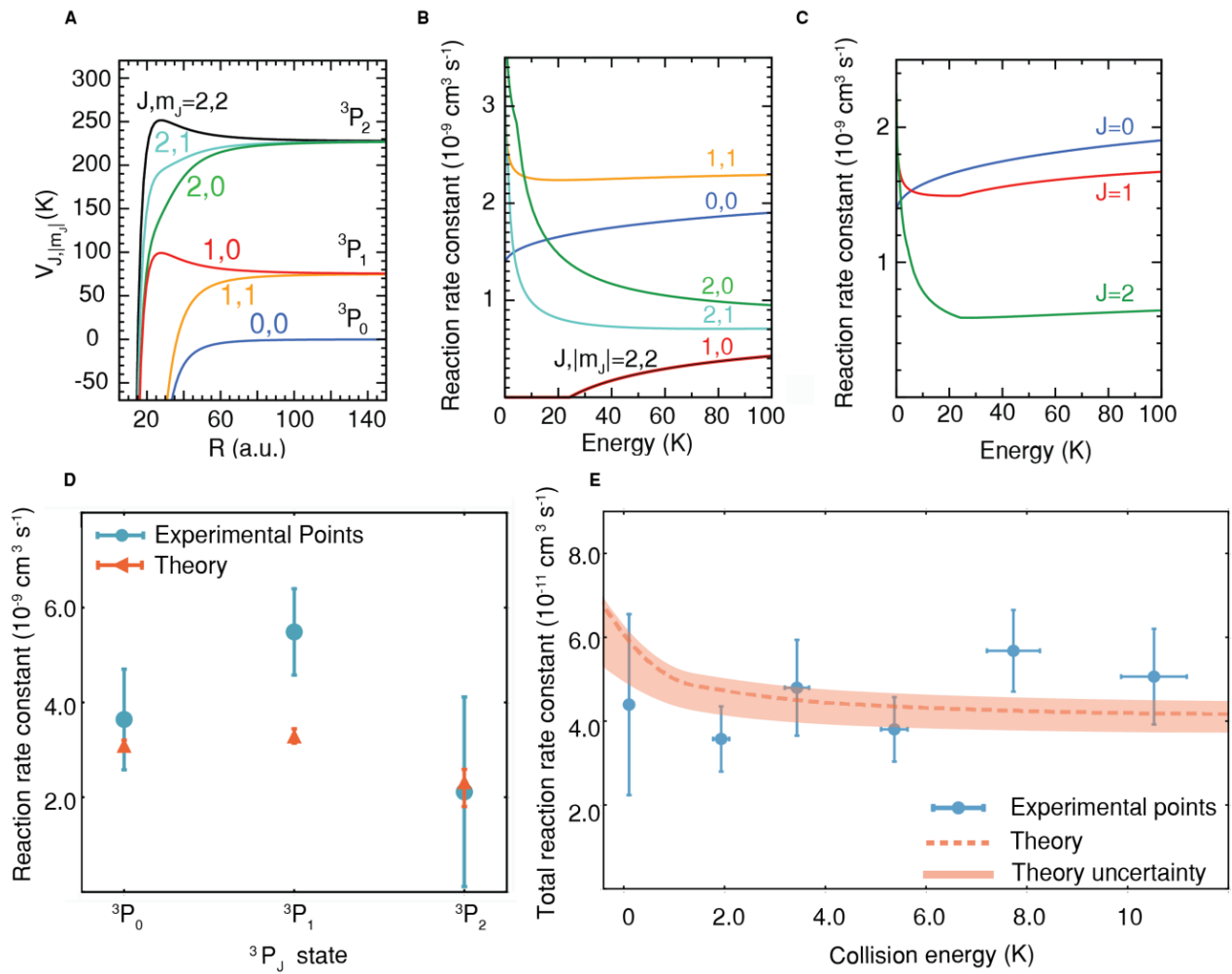



**Figure 4. Individual triplet level molecular potentials and reaction rate constants.**

(**A**) The molecular potential for each triplet sublevel. (**B**) The subsequent energy dependent rate constants obtained from capture theory. (**C**) The $m_J$ averaged rate constants assuming equal population of each $m_J$ level for each J level. (**D**) The rate constant of each individual triplet state, measured by depopulating the other triplet states through optical pumping and acquiring reaction kinetics data. Solutions of differential equations were fitted to approximately 250 kinetic data points to obtain reaction rate constants at each triplet setting, with experimental uncertainties expressed at the one-sigma level *(17)*. Theoretical estimates, along with uncertainty bands associated with the polarizability and quadrupole moment values used to construct the molecular potentials in (A), are presented alongside the data. (**E**) The temperature dependence of the total reaction rate compared to theory by varying the micromotion energy of ions in the LQT and recording reaction kinetics data at each temperature, with the theoretical uncertainty denoted by the thickness of the theory band. Roughly 250 data points were collected at each collision energy, and experimental uncertainties are presented at the one-sigma level.




**Acknowledgements:**

We thank Wes Campbell, Steven Schowalter, Alexander Dunning, and Elizabeth West for insightful conversations. PP would also like to thank Kent Purser for foundational discussions. This work was supported by National Science Foundation (PHY-1205311 and DGE-1650604) and Army Research Office (W911NF-15-1-0121 and W911NF-14-1-0378) grants. All data presented in this work is available through the Harvard Dataverse: [https://dataverse.harvard.edu/dataverse/baoca_2017](https://dataverse.harvard.edu/dataverse/baoca_2017).


**Supplementary Materials:**

Materials and Methods

Supplementary Text

Figures S1-S3

Tables S1-S2



# Supplementary Materials for

## Synthesis of mixed hypermetallic oxide BaOCa$^+$ from laser-cooled reagents in an atom-ion hybrid trap


Prateek Puri[1], Michael Mills[1], Christian Schneider[1], Ionel Simbotin[2], John A. Montgomery, Jr.[2], Robin Côté[2], Arthur G. Suits[3], and Eric R. Hudson[1]

correspondence to: eric.hudson@ucla.edu


**This section includes includes:**





**Materials and Methods**

<u>Experimental design</u>

The hybrid atom-ion trap utilized in this study consists of a co-located Ca MOT and a segmented LQT, which is radially coupled to a time-of-flight mass spectrometer (ToF). The MOT utilizes a standard six-beam geometry, where three pairs of orthogonal, counter-propagating Doppler-cooling beams drive the *4s4p* $^1P_1 \leftarrow 4s^2\,{}^1S_0$ cooling transition in Ca at 423 nm. An additional laser at 672 nm repumps Ca atoms which decay into the metastable *3d4s* $^1D_2$ state back into the cooling cycle by promoting the atoms to the upper *4s5p* $^1P_1$ state, which quickly decays back to the ground state (Fig. 3C). Current coils arranged in the anti-Helmholtz configuration create the quadrupolar magnetic field necessary for the MOT, with a field gradient of ~60 G/cm produced at the magnetic field null near the intersection point of the cooling lasers.

Neutral Ca atoms are introduced into the chamber by a heated getter unit and are subsequently decelerated by far-detuned cooling beams to velocities capturable by the MOT. Imaging of the atoms is performed by two near-orthogonal EMCCD cameras equipped with laser line filters to observe fluorescence from the 423 nm cooling transition.

Throughout the experiment, spatially overlapped 493 nm cooling (*6p* $^2P_{1/2} \leftarrow$ *6s* $^2S_{1/2}$) and 650 nm repump (*6p* $^2P_{1/2} \leftarrow 5d\,{}^2D_{3/2}$) external cavity diode lasers (ECDL) cool the Ba$^+$ ions in both the axial and radial dimensions, while the laser cooled Ba$^+$ ions themselves provide sympathetic cooling for the molecular ions co-trapped in the LQT. Imaging of Ba$^+$ is performed by collecting ion fluorescence from the 650 nm transition with a single reentrant EMCCD camera; prolate spheroid distributions can be fitted to fluorescence profiles of the ions to extract spatial estimates of the crystal size.

The segmented LQT has a field radius $r_0 = 6.85$ mm and electrode radius $r_e = 4.50$ mm. The $\Omega = (2\pi)$ 680 kHz trapping potential is applied asymmetrically with one diagonal pair of electrodes having an rf amplitude $V_{rf} \sim 180$ V and the other pair at rf ground. Axial confinement is provided by biasing the outer segments of the LQT by $V_{ec} \sim$ 4 V. Central electrodes are also biased to compensate for stray DC electric fields that displace the ions from the trap null and cause additional excess micromotion energy. Ba$^+$ is loaded into the LQT by ablating an in-vacuum BaCl$_2$ target with a pulsed 1064 nm Nd: YAG laser. In order to purge the ion trap of heavier species initially loaded from the target, symmetric DC voltages are applied to a pair of diagonally opposed electrodes, thereby increasing the Mathieu a-parameter such that species heavier than Ba$^+$ are ejected from the LQT.

The size of the Ba$^+$ crystal can be further controlled by iteratively lowering the endcap voltages to decrease the axial trap depth of the LQT, resulting in ion loss. Using this technique, Ba$^+$ samples ranging from large three-dimensional Coulomb crystals to single-ion-resolved linear chains are initialized in the LQT. Since the collision energy of ions with the MOT atoms is dominated by the excess micromotion energy of the ions, which scales quadratically with the radial distance from the trap null, control of ion crystal size also affords control of reaction energetics.

Since the relative geometry of the MOT cameras and the ion camera is well known, the spatial overlap between the Ca atoms and the ions can be determined by transforming two-dimensional MOT and ion density distributions obtained from each fluorescence



image into the same coordinate system and integrating over both. First, the ions are imaged in both of the MOT cameras, which are also equipped with laser line 493 nm filters for detecting $Ba^+$ fluorescence. Knowledge of the spatial offset between the MOT and the ion crystal in three orthogonal dimensions is necessary for a complete overlap calculation. An individual MOT camera is typically only capable of detecting spatial offsets between the atoms and ions in the two dimensions spanned in its image plane. However, since we employ two MOT cameras whose image axes are non-parallel, the atom-ion offset in the dimension inaccessible to one camera can be tracked by projecting the offset detected in the other camera onto that dimension, allowing for a full three dimensional determination of the overlap factor as well as spatial profiling of the MOT, which is modeled with a Gaussian density distribution. Further, since the spatial ion distribution changes as a function of MOT immersion time due to charge exchange collisions and other chemical reactions, the time dependency of the overlap can be tracked by continuously monitoring fluorescence images of the ions throughout the experiment.

The overlap factor can be calculated as

$$\hat{O}(t) = \iiint_{-\infty}^{\infty} \prod_{i=1}^{3} e^{-\frac{(x_i - \delta_i)^2}{\omega_i^2}} \bar{n}_{ion}(x_1, x_2, x_3, t) \, dx_1 \, dx_2 \, dx_3$$

where $\hat{O}(t)$ is the time-dependent overlap factor, $x_i$ is the spatial coordinate in the $i^{th}$ dimension, $\delta_i$ is the spatial offset between the ions and the MOT in the $i^{th}$ dimension, $\omega_i$ is the $e^{-1}$ decay length determined from Gaussian fits to MOT fluorescence images in the $i^{th}$ dimension, and $\bar{n}_{ion}(x_1, x_2, x_3, t)$ is the time-dependent, integral-normalized ion density distribution, which satisfies the condition:

$$\iiint_{-\infty}^{\infty} \bar{n}_{ion}(x_1, x_2, x_3, t) \, dx_1 \, dx_2 \, dx_3 = 1.$$

## ToF-MS functionality

Developed in previous work *(21)*, the radially coupled ToF described in this work is utilized for mass-selective species detection within the ion trap. After the LQT has been loaded with ions, the ToF operates by suspending the trapping rf applied to the LQT electrodes for ~1.5 $\mu$s and subsequently applying a high voltage DC signal *(24)* to the trap rods to guide the ions into a field free drift tube, where they eventually register an ion-number dependent signal on a channeltron detector. Further, as noted in recent work, utilizing this method with laser-cooled ion species, as well as with sympathetically cooled co-trapped ions, yields dramatic increases in mass resolution *(25)*. Such techniques were employed in this work to achieve isotopic resolution for all ion trap species of interest.

The relationship between ToF arrival time and mass for a given species is given by $t_a = k_a \sqrt{m_a} + t_0$, where $t_a$ is the arrival time for the species of interest, $m_a$ is the mass of the species of interest, $k_a$ is a proportionality constant, and $t_0$ is an offset time *(40)*. The ToF is calibrated by measuring channeltron arrival times for species of known masses in our system ($BaCl^+$, $Ba^+$, $BaOH^+$, and $BaOCH_3^+$) and performing a fit using the above functional form to determine $k_a$ and $t_0$.



There are two main sources of mass determination error to consider. Firstly, there is an uncertainty in peak arrival time associated with the temporal spread of each species' ToF signal. To this end, a Gaussian fit is applied to each mass signal to extract this spread in arrival time. Secondly, errors in our actual ToF calibration can be determined from the fit errors to the free parameters in the above arrival time equation. In addition to the errors associated with the parameters themselves, the accuracy of our mass determination for unknown species in the trap increases as the mass of these unknown species nears the calibration masses, since the calibration is inherently well known at these points. For example, in the photofragmentation experiment detailed in the manuscript, $Ca^+$ is further away in mass from its nearest calibration mass ($Ba^+$ - 137.9 amu) than $BaO^+$ is ($BaOH^+$ - 154.9 amu), and while both species are still isotoptically resolvable, the associated calibration-related mass determination error is greater for the former. This calibration-based error is calculated from spreads in 95% confidence interval bands applied to our calibration fit, assuming ion arrival time uncertainties as determined by the above-mentioned Gaussian mass signal fit. By adding our confidence interval uncertainty in quadrature with the mass error determined from propagating error through the arrival time equation, we are able to extract conservative overall mass determination errors, and these values are reported alongside all experimentally measured masses presented in the manuscript.

Calcium electronic state population modeling and manipulation

A recent study *(28)* utilized relativistic many body theory to evaluate electronic dipole transition rates for both spin-forbidden and spin-allowed transitions within the first 75 energy levels of Ca. Using these values, a comprehensive rate equation model for Ca can be constructed that considers natural spontaneous emission as well as stimulated emission and absorption caused by near-resonant laser fields present in the MOT trapping volume. A representative differential equation for state $i$ with a monochromatic laser driving a transition from state $i$ to $k$ is given by

$$\frac{dN_i}{dt} = \sum_{j>i} \Gamma_{ji} N_j - \sum_{j<i} \Gamma_{ij} N_i - \frac{N_i}{\tau_{Loss}} + \Gamma_{ki} \left(\frac{\pi^2 c^3}{\hbar \omega_{ik}^3}\right) \frac{I_{ik}}{2\pi c} \frac{\Gamma_k}{(\omega_{ik} - \omega_l)^2 + \left(\frac{\Gamma_k}{2}\right)^2} (N_k$$
$$- \frac{2j_k + 1}{2j_i + 1} N_i)$$

where $N_i$ is the amount of population in the $i^{th}$ state, $\Gamma_{ij}$ is the decay rate of state $i$ to $j$, $\tau_{Loss}$ is the time in which an uncooled atom drifts outside of the MOT trapping region (for our parameters, this value is ~1.7 ms for the *4s4p* $^3P_0$ and $^3P_2$ states and $\infty$ otherwise), $c$ is the speed of light in vacuum, $\hbar$ is the reduced Planck constant, $\omega_{ik}$ is the angular transition frequency between state $i$ and $k$, $\omega_l$ ($I_{ki}$) is the angular frequency (intensity) of the applied laser, $\Gamma_k$ is the natural linewidth of state $k$, and $j_i$ is the total angular momentum quantum number of state $i$. By incorporating experimental laser detunings and intensities into the



rate model, accurate electronic state populations can be calculated for each energy level in Ca.

While the metastable state populations in Ca are difficult to detect directly due to low atom number densities, experimental measurements of population in the *4s4p* $^1P_1$ level involved in the cooling transition are possible through standard absorption and fluorescence imaging techniques. Such measurements agree with predictions made by the rate model, and, when further combined with the excellent agreement between theory and experiment presented elsewhere *(29)*, demonstrate that the rate equation model is an accurate predictor of Ca electronic state populations.

There are several transitions utilized in this study to control electronic state populations, and Table S2 summarizes the key properties of each transition along with the laser sources used to drive them.

The primary method for varying population in the $^3P_J$ levels is controlling the power of the 423 nm beam, which is used to drive the cooling transition for the MOT. The 423 nm beam intensity directly affects the amount of population in the $^1P_1$ state and, since the $^3P_J$ levels are populated primarily through decay channels from the $^1P_1$ state, also has an effect on the total $^3P_J$ population.

The 429 nm, 431 nm, and 446 nm beams are used to depopulate specific $^3P_J$ levels in order to isolate population in an individual triplet state to measure its reaction rate, while the 672 nm and 411 nm beams each drive separate repump transitions for the MOT. Since the 411 nm beam serves as a more efficient repump than the 672 nm beam, and consequently more effectively prevents population buildup in the dark $^3P_J$ levels, switching between these repumps allows for further control of the total $^3P_J$ population.

Kinetic data reaction rate extraction

In order to probe the time evolution of the reaction, ions are pulsed into the ToF after a variable amount of interaction time with the Ca MOT, allowing the amount of each ion in the LQT to be monitored as a function of reaction time. Each ion trap species can potentially be subject to the following decay and production processes:

1) Charge exchange with neutral calcium
2) Production and consumption due to chemical reactions with either $CH_3OH$ or Ca
3) Photodissociation from cooling beam lasers and natural trap loss

A sample differential equation for the time evolution of ion number in the trap is given by

$$\frac{dN_i}{dt} = -n_{Ca}\hat{O}_{Ca}(t)\,k_{i,CEX} + \sum_{\alpha,j} n_\alpha \hat{O}_\alpha(t)\,k_{\alpha+j\rightarrow i}\,N_j - \sum_{\alpha,j} n_\alpha \hat{O}_\alpha(t)\,k_{\alpha+i\rightarrow j}\,N_i - \lambda_L N_i$$

where $N_i$ is the amount of the $i^{th}$ ion present, $n_\alpha$ is the density for neutral reactant $\alpha$, $\hat{O}_\alpha(t)$ is the time-dependent overlap factor between the ions and neutral reactant $\alpha$, as defined earlier, $k_{i,CEX}$ is the charge exchange reaction rate for the $i^{th}$ ion, $k_{\alpha+j\rightarrow i}$ is the reaction rate



between neutral reactant $\alpha$ and ion $j$ to form ion $i$, and $\lambda_L$ is the loss term accounting for both photodissociation and natural trap loss in the LQT. A coupled set of differential equations for the $Ba^+$, $BaOCH_3^+$, $BaOH^+$, and $BaOCa^+$ ions is numerically solved and fitted to the kinetic ToF data in order to extract values for the unknown parameters, with a typical set of kinetics data and corresponding fitted solutions shown in the inset of Fig 2C. While every species experiences charge exchange with Ca or natural trap loss, the only chemical reactions observed to have significant reaction rates are

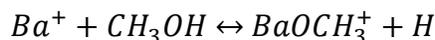

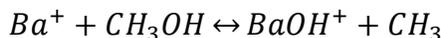

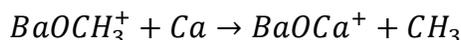

and thus, for computational convenience, only these reactions are considered when constructing the system of differential equations. After numerically solving the above mentioned system, values for $k_{Ca+BaOCH_3^+ \rightarrow BaOCa^+}$ (abbreviated as $k_t$ in the Main Text) can be extracted.

Since the initial amount of $Ba^+$ determines both the amount of $BaOCH_3^+$ reagent that will accumulate in the trap prior to MOT immersion and, consequently, also the eventual amount of $BaOCa^+$ produced, slight fluctuations in ion initialization due to ablation loading variability can introduce additional noise to the kinetics data. In order to account for these fluctuations, prior to fitting, each ion amount is normalized by the initial amount of $Ba^+$, which can be readily obtained from either fluorescence imaging or ToF detection.

Electronic Structure Calculations

To aid in the interpretation of the experimental results, electronic structure calculations were performed for the Ca + $BaOCH_3^+$ → $BaOCa^+$ + $CH_3$ reaction. Optimized geometries for $BaOCH_3^+$ and $BaOCa^+$ and their fragments were obtained from density functional theory (DFT) using the triple-zeta correlation consistent basis sets (cc-pwCVTZ on calcium and barium and cc-pVTZ on hydrogen, carbon, and oxygen) and the B3LYP density functional. The inner shell electrons of calcium and barium were described by an effective core potential (ECP). Harmonic vibrational frequencies were computed at all stationary points. The calculated vibrational frequencies were used to characterize the stationary points as minima or saddle points and to obtain vibrational zero point energies. Coupled cluster theory including single and double excitations with perturbative triples, denoted CCSD(T), was used to estimate thermochemical energy differences. To check the validity of the DFT geometries for this problem, the CCSD(T) energies of the stationary points were recalculated at geometries obtained from second-order Møller-Plesset (MP2) theory, and the changes in thermochemical energy differences were less than 1 kcal/mol. DFT and MP2 offer different approaches to the electron correlation problem, but they predict geometries of generally comparable accuracy. Discrepancies between them would



be an indication that a higher level of theory should be used, but their agreement here suggests such methods are not warranted. The electronic structure calculations were performed using the Gaussian 09 and Molpro 2012 program packages *(24, 25)*.

The calculated results show the Ca ($^1S_0$) + BaOCH$_3^+$ $\rightarrow$ BaOCa$^+$ + CH$_3$ reaction to be exothermic by 5.3 kcal/mol at the CCSD(T)/cc-pVTZ level of theory. Interestingly, at this level of theory, most of the exothermicity results from a loss of vibrational zero point energy between reactants and products. At the more expensive CCSD(T)/cc-pV5Z level of theory, the heat of reaction is increased to 8.4 kcal/mol.

The dipole moment calculation was performed at the B3LYP/cc-pVTZ level, using broken symmetry unrestricted Hartree-Fock (UHF) for the singlet neutral, whereas the singlet-triplet splitting was calculated at the projected third-order Møller-Plesset (PMP3/cc-pVTZ) level of theory, again using broken symmetry UHF for the singlet neutral. Further, the ionization energy was determined at the CCSD(T) with UHF reference/cc-pVTZ level.

The intrinsic reaction coordinate (IRC) calculation, depicted in Fig. S1, was performed at the B3LYP/cc-pVTZ level of theory, revealing the existence of two bound BaOCH$_3$Ca$^+$ complexes, one in the entrance channel and one in the exit channel. The structure and relative energy of both of these structures are investigated at the CCSD(T)/cc-pVTZ level of theory, indicating the existence of a barrier to the reaction, whose height is 10.2 kcal/mol. The IRC calculations were performed using the Hessian based predictor-corrector integrator method of Hratchian and Schlegel. The Hessian was recomputed at each integration step.

It is possible that multi-reference effects may be significant in this system. To assess the need for a multi-reference description, valence complete active space (CAS) multi-configurational self-consistent field (MCSCF) calculations were performed on all the singlet and triplet structures presented in Fig. 2A. The weight, defined as the square of the configuration interaction (CI) coefficient, of the reference Hartree-Fock configuration is shown to be 94% in the triplet state and 80% in the singlet for the transition state (TS) structure, indicating that multi-reference effects play a role in the singlet, but not the triplet, TS. The second most important configuration in the singlet TS has a weight of 13%, while all others are found to be less than 2%. Natural orbital analysis of the broken symmetry UHF wavefunction for the singlet TS shows two natural orbitals with significant fractional occupation (1.2 and 0.8), indicating the broken symmetry UHF gives an approximate description of the two most important configurations in the CAS wavefunction. A CCSD(T)/VTZ calculation using the broken symmetry UHF wavefunction as a reference results in a 3.4 kcal/mol increase in the barrier height. Further, CI coefficients for the remainder of the structures along the triplet pathway demonstrate reference configuration weights $\gtrsim 93\%$, allowing us to conclude that multi-reference effects would not significantly alter the conclusions of our computational study (Table S1).

Description of the magnetic trap and the experimental sequence of the shuttling experiment

Ca atoms in the $|^3P_2, m_J = 2\rangle$ are trapped in the magnetic field gradient produced by the anti-Helmholtz configuration of the MOT coils (Fig. 3A). Due to the linear



dependency of the trapping potential on the magnetic quantum number and the Landé g-factor, only $|^3P_2, m_J = 1, 2\rangle$, $|^3P_1, m_J = 1\rangle$, $|^1P_1, m_J = 1\rangle$, and $|^1D_2, m_J = 1, 2\rangle$ atoms have stable trapping forces. However, since the last three states have very short radiative lifetimes (< 2 ms) compared to that of the $^3P_2$ states (~118 min), after only a few milliseconds, the magnetic trap contains purely $|^3P_2, m_J = 1, 2\rangle$ atoms. Further, due to the smaller potential depth of the $|^3P_2, m_J = 1\rangle$ atoms, the atomic density of the magnetically trapped $|^3P_2, m_J = 2\rangle$ atoms is ~10x larger, resulting in a magnetic trap dominated by $|^3P_2, m_J = 2\rangle$ Ca atoms. We note that $m_J$ here is defined with respect to the trap magnetic field direction, while the relative velocity vector defining the reaction is isotropically distributed, meaning the Ca $m_J$ sublevel is not controlled along the reaction quantization axis.

In the magnetic trapping experiment, ions are first initialized as described earlier. To ensure that the reaction can only occur between the magnetically trapped Ca atoms and BaOCH$_3^+$ molecules, the voltages of the LQT are adjusted so the BaOCH$_3^+$ ions are first displaced from the center position of the MOT by ~3 mm, precluding any background reactions from direct MOT-BaOCH$_3^+$ overlap. After the magnetic trap is loaded to capacity in ~1 s, the MOT cooling beams are extinguished. The magnetic trap, which is not laser dependent, has a lifetime of ~500 ms, while the MOT lifetime after beam shuttering is only ~1 ms. After waiting 50 ms to completely purge the MOT atoms from the magnetic trap volume, the LQT voltages are adjusted to shuttle BaOCH$_3^+$ ions into the center position of the magnetic trap, allowing them to react directly with a nearly pure sample of $|^3P_2, m_J = 2\rangle$ atoms, as shown in Fig 3A. After reacting for ~ 2 s, at which point the magnetic trap is nearly depleted, the ions are again displaced from the center of the magnetic trap, the MOT beams are unshuttered to reload the magnetic trap, and the process is repeated. After a variable amount of these cycles (up to ~100), the LQT products are pulsed in the ToF, yielding mass spectra that can be analyzed for BaOCa$^+$ production. To further prove this reaction is driven by the trapped $|^3P_2, m_J = 2\rangle$ atoms, a control case is also presented. In this case, a frequency-doubled titanium-sapphire laser is used to optically pump population out of the $^3P_2$ state via the 446 nm $4s4d$ $^3D_2 \leftarrow 4s4p$ $^3P_2$ transition throughout the experiment, preventing any metastable Ca atoms from remaining in the magnetic trap.

An atom number density is required to calculate reaction rates from these experiments. While absorption imaging methods are typically used to assess the density of atom traps, in this case, the magnetic trap is not dense enough for this technique to yield resolvable values and alternative methods need to be pursued.

Determination of an atom number density requires information both on the total number of atoms inside of the magnetic trap and on the spatial distribution of these atoms. In order to measure the trapped atom number, optical pumping was utilized to transfer the magnetically trapped triplet atoms into the ground state. At this point, the MOT cooling beams were immediately introduced into the chamber for ~5 ms and fluorescence from the magnetically trapped atoms was collected. Fluorescence from a small amount of atoms loaded into the MOT from the getter source was also recorded during this interval; however, since the MOT loading time is ~100 ms, this background is negligible and can further easily be accounted for through background subtraction. Using a calibration of fluorescence to atom number based on observing a Ca MOT of known density, magnetic



trap total atom numbers were then calculated. However, no spatial information about the magnetic trap was offered by this measurement since the trap was optically depopulated before fluorescence collection.

The spatial distribution of the magnetic trap was theoretically estimated by using measured magnetic fields and the Landé g-factors, $g_F$, of the Ca triplet atoms to calculate the potential of the magnetic trap, which is given by $U_B = \overrightarrow{\mu_M} \cdot \vec{B}$, where $U_B$ is the magnetic trap potential energy, $\vec{B}$ is the trap magnetic field and $\overrightarrow{\mu_M}$, the magnetic moment, is given by $g_F \mu_B \vec{J}$, where $\mu_B$ is the Bohr magneton, and $\vec{J}$ is the total angular momentum vector. Combining the atom number measurement with a calculation of the spatial distribution of the magnetic trap derived from the trap potential, trap densities of ~4 x $10^7$ atoms/cm$^{-3}$ were obtained. However, uncertainties in factors such as the optical pumping transfer efficiency and unaddressed effects in the theoretical model, such as the effect of stray magnetic fields on the trapping potential, limit the accuracy of this estimate. While uncertainties due to these effects are difficult to experimentally quantify, we expect an overall error of ~5x in our density estimate.

After the magnetic trap density was estimated, reaction rate values were obtained by fitting the solutions of the reaction kinetics differential equations to the shuttling experiment data (see above section on kinetic data reaction rate extraction), with the order of magnitude accuracy reported in the main text set primarily by the uncertainty in magnetic trap density.

Long-range capture theory and Langevin rate calculation

Given that calculations suggest the reaction is barrierless, we expect that the observed reaction rate can be estimated from long-range capture theory. As the ion approaches the Ca atom, the quadrupole moment of the Ca $^3P_J$ state leads to a long-range R$^{-3}$ interaction, in addition to the usual R$^{-4}$ polarization potential. We assume that both interactions take place at large separations between reactants, and treat BaOCH$_3^+$ as a point charge. Using treatment developed in other work *(33)*, we evaluate the potential curves for Ca ($^3P_J$) in states (J, $|m_J|$), using the $^3P_2$ quadrupole moment Q and the static polarizabilities $\alpha_{xx}$ and $\alpha_{zz}$ calculated elsewhere *(32)*. Accounting for the spin-orbit coupling *(33)* and using the fine structure intervals from NIST *(34)*, we obtain the potential curves shown in Fig. 4B. The curves for $\pm m_J$ are identical, resulting in three distinct curves for J=2, two for J=1, and a single curve for J=0. The effect of the quadrupole is non-trivial, leading to barriers that reduce reaction rates for some channels or more attractive curves that increase the reaction rates for others. We assume a conservative estimated variation of 10% of the adopted values (in atomic units) of Q = 12.9, $\alpha_{xx}$ = 295.3, and $\alpha_{zz}$ = -28.37 from *(35)* to account for the range of published values for these quantities *(37-39)*.

To compute theoretical energy-dependent reaction rates, we employ a simple Langevin capture model *(35, 36)*. The cross section is given by σ = πb$^2$, where the impact parameter $b = \hbar \, (\ell + \frac{1}{2})/p$ is given by the maximum angular momentum $\ell$ allowing the atom-ion pair to reach short separation for a given relative collision energy $E = \frac{p^2}{2\mu}$, where $\mu$ is the reduced mass of the system and $p$ is the linear momentum. An energy dependent rate constant k = vσ is obtained by multiplying σ with the relative velocity $v = p/\mu$.



## Collision energy spatial averaging

The trajectory of a trapped ion in in the low Mathieu-q limit can be approximated (*41*) as

$$r(t) = r_0 \left(1 + \frac{q}{2}\cos(\Omega\, t)\right)\cos\left(\frac{q}{2\sqrt{2}}\Omega\, t\right)$$

where $r(t)$ is the ion radial coordinate at time $t$, $r_0$ is the ion's initial radial displacement, and $q$ is the Mathieu-q parameter, given as $q = \frac{2eV_{RF}}{m\, r_0^2 \Omega^2}$, where $e$ is the electronic charge and $m$ is the mass of the ion (see above *Experimental design* section for trap parameter definitions and values).

This motion consists of harmonic oscillation at a secular trap frequency, $\frac{q}{2\sqrt{2}}\Omega$, modulated by faster micromotion oscillation at the rf drive frequency of the ion trap. Under laser-cooled conditions, the secular portion of the ion's motion is cooled to $\leq 1$ mK. This energy is negligible relative to the micromotion energy of the ion (*42*), and consequentially, the ion velocity can be modeled as follows:

$$v(r_0, t) = -r_0 \Omega \frac{q}{2}\sin(\Omega\, t)$$

The velocity probability distribution function is then given by the probability distribution function for a classical harmonic oscillator at frequency $\Omega$ as

$$P(r_0, v) = \begin{cases} 0 & \left(\frac{q\Omega}{2}r_0\right)^2 < v^2 \\ \dfrac{1}{\pi}\sqrt{\dfrac{1}{\left(\frac{q\Omega}{2}r_0\right)^2 - v^2}}, & \left(\frac{q\Omega}{2}r_0\right)^2 \geq v^2 \end{cases}$$

This velocity probability function, which depends on the initial ion radial position, can further be weighted by the ion spatial density distribution as

$$P_V(v) = \int \bar{n}_{ion}(r, z, \phi) P(r, v)\, d^3 r$$

where $P_V(v)$ is the ion velocity probability distribution and $\bar{n}_{ion}(r, z, \phi)$ is the ion-density distribution expressed in cylindrical coordinates, whose parameters are characterized through fits to fluorescence images obtained by our EMCCD camera. After obtaining the velocity probability distribution, it is then straightforward to calculate $\bar{E}_{col}$, the average collision energy, as $\bar{E}_{col} = \int_{-\infty}^{+\infty} P_V(v)\frac{1}{2}\mu\, v^2 dv$, which is the value reported in the main text in Fig. 4D, where the horizontal error bar for each point stems from slight variations in initial ion amount, and hence ion-density distribution, at each collision energy setting. The variations in initial ion amount can be monitored through fluorescence imaging and can further be quantified through ToF analysis, whereby an initial amount of Ba$^+$ can be inferred from knowledge of the final amount of Ba$^+$ collected in the ToF spectrum and the amount of interaction time with reactive neutrals (see above *Kinetic data reaction rate extraction* section)."

Statistical Analysis: Exclusion of other electronic states from participation in the reaction

Before the $^3P_J$ reaction pathway for BaOCa$^+$ was identified, it was unclear which electronic state was participating in the reaction, and statistical methods were employed to exclude alternative reaction pathways involving other states populated during Ca laser cooling, i.e. the $4s^2$ $^1S_0$, $4s4p$ $^1P_1$, $4s5p$ $^1P_1$ and $3d4s$ $^1D_2$ states.

First, laser parameters were manipulated in order to vary each electronic state's population over a suitable range, as described earlier. Reaction rates were then extracted for the different population settings and plotted as a function of individual state population. Since the proposed reaction involves a single Ca atom, if a particular electronic state is involved in the reaction, a linear relationship between state population and production rate is expected, and consequently, a linear fit is applied to each data set (Fig. S2, A-E).

In order to assess how well the observed reaction rate data matched the expected linear dependency on state population, the reduced chi-squared statistic, $\chi^2_{red} = \frac{1}{n-f} \sum_i \frac{(M_i - C_i)^2}{\sigma_i^2}$, was calculated for each model, where $n$ is the number of observations, $f$ is the number of fit parameters, $M_i$ is the model's prediction for the $i^{th}$ data point, $C_i$ is the average value for the $i^{th}$ data point, and $\sigma_i$ is the standard error for the $i^{th}$ data point.

According to standard $\chi^2_{red}$ analysis, the model that best represents the data is the $^3P_J$ production pathway, yielding a $\chi^2_{red} = .22$ (Fig. S2E). The $4s5p$ $^1P_1$ pathway offers the next lowest $\chi^2_{red}$ value; however, BaOCa$^+$ production is still observed when the 672 nm repump is shuttered and the $4s5p$ $^1P_1$ state is unpopulated (Fig. 3C), thus invalidating this pathway as the sole means of BaOCa$^+$ production.

Further, the vertical intercept of the $^3P_J$ fit is consistent with zero, which suggests that there are no other competing reaction entrance channels. For example, if non-adiabatic coupling from excited singlet surfaces to other electronic states were occurring and resulting in reaction, and the populations of such sates were not correlated with triplet population, we should observe a background reaction rate that would exist even with a 0% triplet population, which is what the y-intercept of the fit of Fig. S2E signifies. While we cannot rule out the possibility of such effects occurring, we can experimentally restrict the total reaction rate constant of such events to be $\lesssim 10^{-11}$ cm$^3$/s, which is derived from the $2\sigma$ error in the y-intercept fit and is an order of magnitude smaller than the rates typically accessed in this work.

However, since the population of certain electronic states are coupled to others through spontaneous emission and resonant laser fields, the population of certain singlet channels, such as the $3d4s$ $^1D_2$ state, are correlated with the population of the triplet state. While no other state beside the $^3P_J$ levels demonstrated a linear correlation with BaOCa$^+$ production, this does not necessarily imply no other states are reactive. For example, if an excited singlet state were indeed reactive through non-adiabatic processes, it is possible that enhanced BaOCa$^+$ production due to increasing population in this state could be masked by a corresponding reduction in the population of the triplet state. While we cannot rule out such multi-component reaction pathways from taking place, the above $\chi^2_{red}$ analysis clearly identifies the $^3P_J$ surface as the dominant reaction entrance channel. Moreover, the results of magnetic trap experiment, where the reaction rate between a sample of pure triplet atoms and BaOCH$_3^+$ ions was measured and shown to be consistent with all other observed reaction rates in the study, further suggests that BaOCa$^+$ formation initiates predominantly along the $^3P_J$ Ca potential surface, with all other entrance channels likely negligible on experimental timescales.

**Supplementary Text**

<u>Photodissociation fragment detection</u>

      Although ToF spectra indicated the product of the observed reaction possessed a mass-to-charge ratio consistent with that of $BaOCa^+$, photofragmentation analysis was conducted to further verify the identity of the species. Photodissociation of $BaOCa^+$ was first measured as a function of laser intensity at both ~493 nm and ~422 nm wavelengths (Fig. S3, A and B), which were conveniently accessible with our titanium-sapphire laser system. During data acquisition, $BaOCH_3^+$ ions were first immersed in the MOT for a fixed amount of time to initialize a sample of $BaOCa^+$. Subsequently, photodissociating lasers were introduced into the LQT for varying amounts of time before ejecting the ion trap species into the ToF. A simple exponential decay model $N(t) = N_0 e^{-\Gamma_{PD} t}$, where $N$ is the amount of $BaOCa^+$ present in the LQT, $t$ is the amount of dissociation time, and $\Gamma_{PD}$ is the decay rate of $BaOCa^+$ due to photodissociation, was fit to the data, allowing a dissociation rate to be extracted. This process was repeated over a wide range of laser intensities at each laser wavelength.

      While *ab initio* calculations of the molecular structure of $BaOCa^+$ indicate the energy of the various dissociation asymptotes, estimating the wavelength dependency of the photodissociation cross section is difficult due to lack of information on the excited state potential surfaces of the molecule and their Franck Condon factors with levels in the electronic ground state potential. While neither a 493 nm photon nor a 422 nm photon possesses enough energy to drive to the lowest energy $Ca^+ + BaO$ dissociation asymptote, a two-photon process at either of these wavelengths would have enough energy to surpass the $Ca^+ + BaOCH_3$, $BaO^+ + BaOCH_3$, and the $Ba^+ + CaO$ asymptotes (Fig. S3C), indicating that two-photon photodissociation may be possible at these wavelengths. However, the linear dependency of the photofragmentation rate on intensity (Fig. S3, A and B) is suggestive of a one-photon process, perhaps indicating the first step in the two-photon process was saturated by either the ion laser cooling beams or the dissociating beam itself.

      Regardless of the specifics of the dissociation pathway, identification of any of the $Ca^+$, $BaO^+$, or $Ba^+$ photodissociation fragments would provide further evidence confirming the elemental composition of the molecule, and thus, such a search was conducted. While detection of $Ba^+$ fragments is precluded by the unavoidable background of $Ba^+$ from the initial trapped ion crystal, observation of the $Ca^+$ and $BaO^+$ dissociation products is possible, although difficult due to competing introduction of these species from other processes in the LQT. Namely, $Ca^+$ ion production also occurs through ionization of MOT atoms by the laser cooling beams, and $BaO^+$, whose isotopic signatures may overlap with $BaOH^+$ in the mass spectra, is produced through reactions of the $Ba^+$ crystal with the introduced methanol vapor. In order to separate the background ions from the photofragmentation ions, we employed the well-known LQT mass filtering techniques of

secular excitation introduction and stability region manipulation to selectively remove the unwanted background species.

For $Ca^+$ detection, during MOT immersion, the Mathieu q-parameter was modified such that $Ca^+$ ions were unstable in the ion trap during $BaOCa^+$ production, ensuring no MOT-produced $Ca^+$ ions would remain trapped. After the MOT beams were shuttered, the q-parameter was modified to allow stable trapping of $Ca^+$ ions, at which point the 423 nm photofragmentation laser was introduced into the LQT before the ions were ejected into the ToF. Substantial $Ca^+$ is observed when the photodissociating laser is present (Fig. 1D). When compared to a control case in which the above process is repeated while the photofragmentation laser remained shuttered, the difference in $Ca^+$ production, averaged across ~50 data points, was significant at the $3\sigma$ level, with any background $Ca^+$ in the control case attributable to $BaOCa^+$ dissociation from the ion laser cooling beams.

In order to detect the $BaO^+$ photodissociation fragments, a mass-specific secular excitation frequency was applied during MOT immersion to purge the LQT of $BaOH^+$ molecules formed from methanol reactions with the $Ba^+$ crystal, preventing potential ToF spectrum overlap with $BaO^+$ ions produced from photodissociation. After shuttering the MOT beams, the secular excitation was removed and a 493 nm dissociation beam was introduced into the chamber before ejecting the ions into the ToF to obtain a mass spectrum (Fig. 1D). After averaging across ~50 data points, the difference in $BaO^+$ production compared to the control case was found to be significant at the $3\sigma$ level.

The detection of both $Ca^+$ and $BaO^+$ photodissociation fragments, along with the identification of a mass-to-charge ratio *m/z* consistent with 193.9 amu for the product molecule through ToF analysis, provides strong evidence that the product of the observed reaction is in fact $BaOCa^+$.

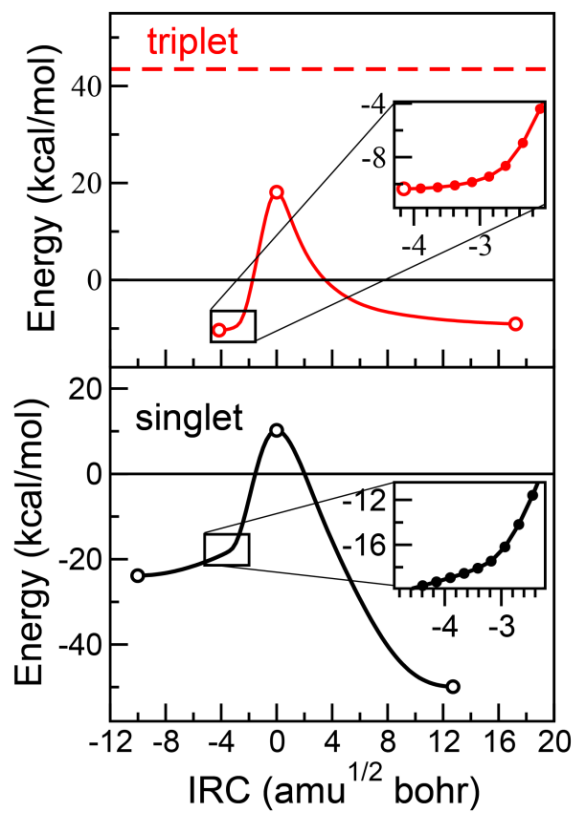

**Figure S1. IRC calculation along the singlet and triplet surfaces**. Energy along the IRC for both the singlet and triplet surfaces calculated at the B3LYP/cc-pVTZ level of theory. The open circles correspond to the stationary points in Fig. 2A, and all energies are given with respect to the singlet ground state reactants, with the initial amount of energy associated with each Ca entrance channel also provided in each graph. The insets display zoomed in views of select regions of the IRC for clarity, where the closed circles indicate calculated energies and the solid lines represent an interpolation.

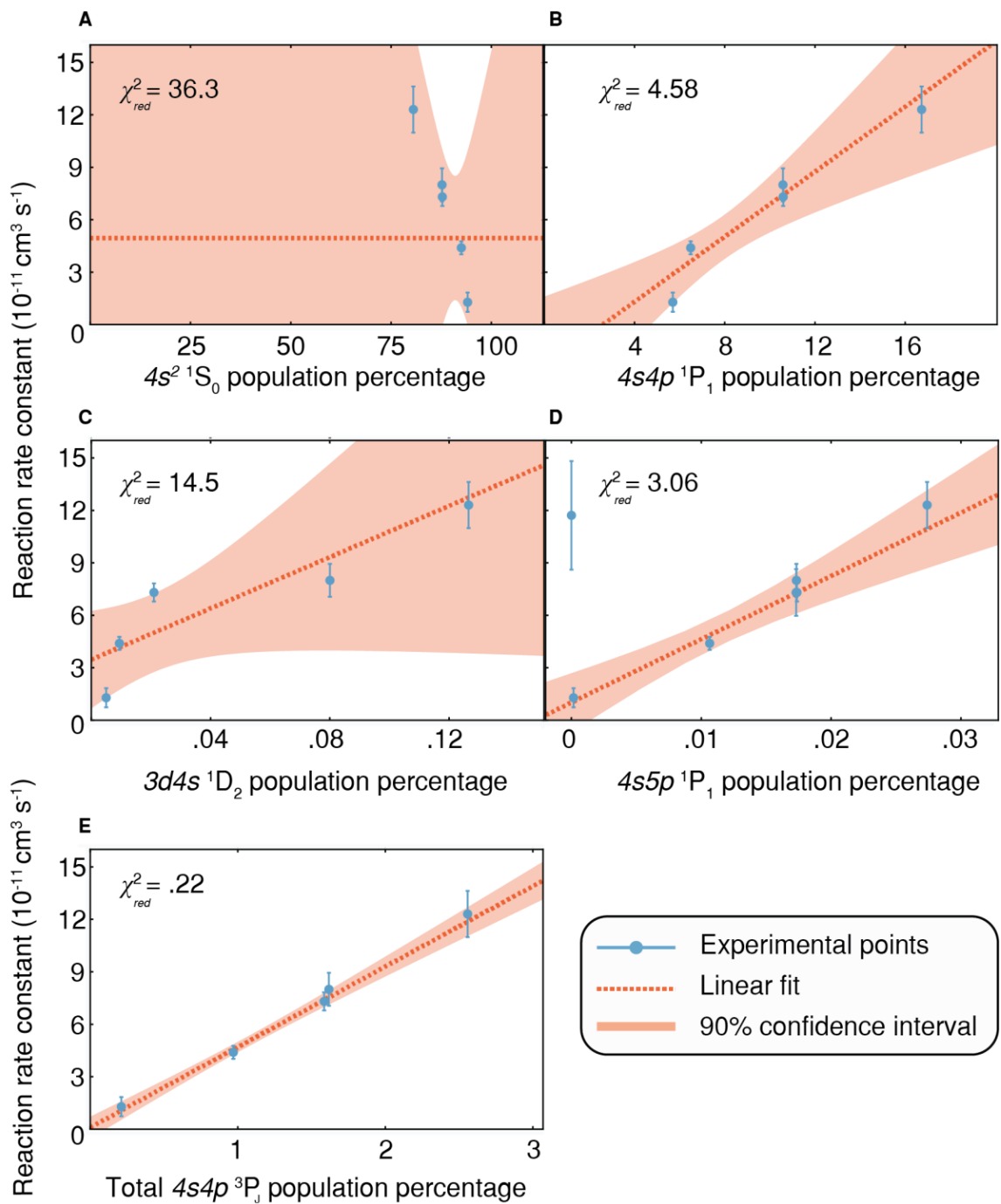

**Figure S2. BaOCa$^+$ production rate dependency on Ca electronic state populations**. Experimentally observed reaction rates are plotted against the **(A)** *4s$^2$* $^1$S$_0$, **(B)** *4s4p* $^1$P$_1$, **(C)** *3d4s* $^1$D$_2$, **(D)** *4s5p* $^1$P$_1$, **(E)** *4s4p* $^3$P$_J$ electronic state populations, which are the only states populated significantly during the Ca laser cooling process. For each plot, experimental points, along with their associated one-sigma uncertainties, are presented alongside linear fits (dashed) and their 90% confidence interval bands. Further, the $\chi^2_{red}$ statistic for each fit is also displayed in the upper left hand corner of the plot. As can be seen from the $\chi^2_{red}$ statistics, the $^3$P$_J$ model best represents the data, supporting the theoretical and experimental findings detailed throughout this report.

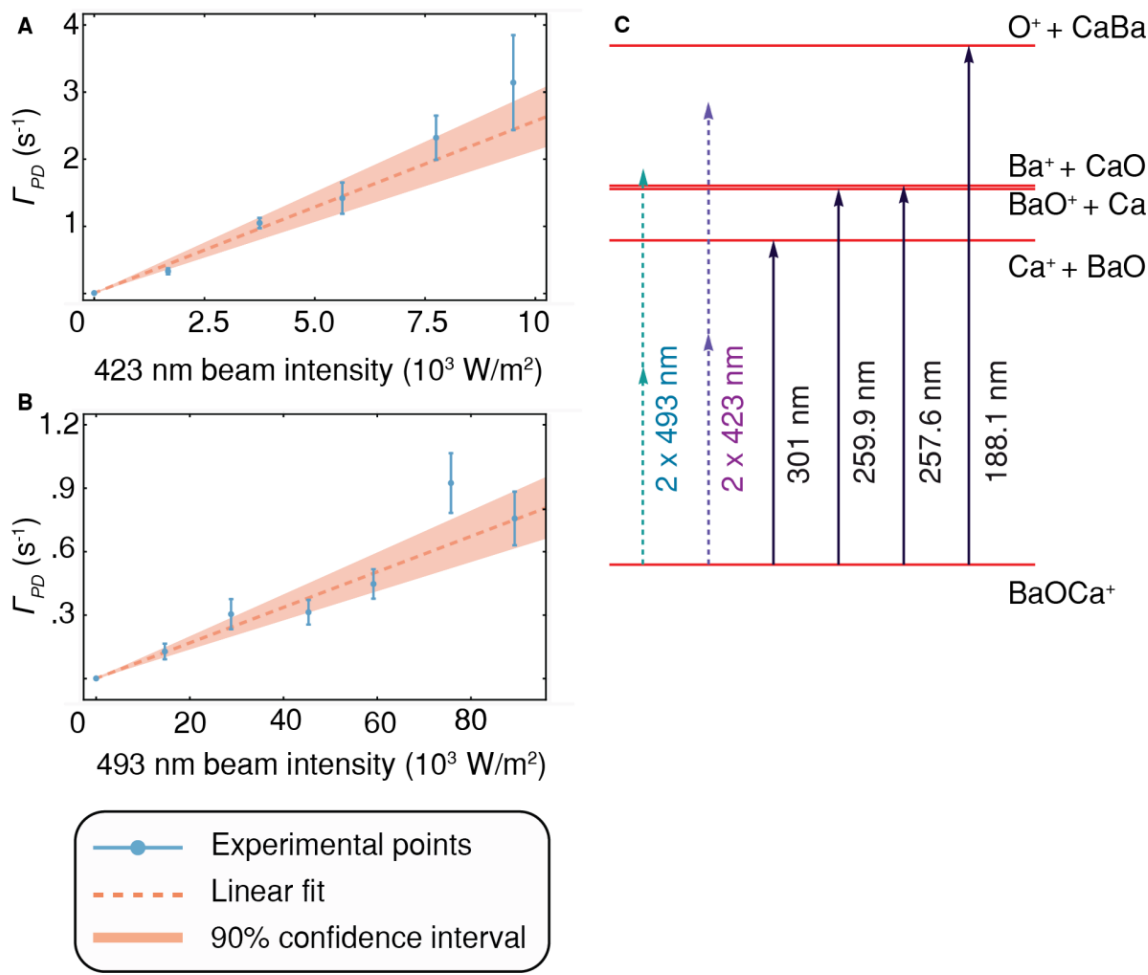

**Figure S3. BaOCa⁺ photofragmentation analysis.**
Experimentally observed photodissociation rates are presented as a function of dissociation beam intensity for both **(A)** 423 nm and **(B)** 493 nm lasers. Roughly 30 data points were acquired at each intensity setting and experimental uncertainties are expressed at the one-sigma level. **(C)** A graphical representation of the various dissociation limits of the molecule, with the asymptotic energy of each represented by arrows labeled in units of photon energy. For comparison, the energy of a two photon process at both 423 nm and 493 nm wavelengths (dashed) is also presented, with both processes possessing an energy above several dissociation limits of the molecule.

| Stationary point | Singlet surface configuration weights (%) | Triplet surface configuration weights (%) |
|---|---|---|
| Ca | 100 (0) | 100 (0) |
| BaOCH$_3^+$ | 94.8 (0.5) | 94.8 (0.5) |
| Entrance channel complex | 88.7 (7.3, 1.6) | 95.8 (0.6) |
| Transition state | 80.3 (13.1, 1.8) | 94.5 (0.5) |
| Exit channel complex | 93.9 (0.8) | 94.8 (0.5) |
| BaOCa$^+$ | 99.2 (0.7) | 99.2 (0.7) |
| CH$_3$ | 96.7 (0.4) | 96.7 (0.4) |

**Table S1. MCSCF calculations to assess multi-reference effects**. CAS MCSCF calculations were performed on all singlet and triplet structures displayed in Fig. 2A to verify multi-reference effects do not play a significant role in our computational study. The weight, as given by the square of the CI coefficient, of the Hartree-Fock configuration is presented for the triplet and singlet structures, along with the weights of the second most significant configuration in parenthesis. For select cases along the singlet surface where the weight of the Hartree-Fock configuration is shown to be less than 90%, the weight of the third most significant reference configuration is also subsequently presented in parenthesis, with all such configurations having weights less than two percent. While secondary configurations appear to be most significant in the singlet transition state, a CCSD(T)/VTZ calculation involving a broken symmetry UHF wavefunction incorporating the two most significant reference configurations indicates a barrier height increase of 3.4 kcal/mol, further confirming the existence of a barrier to reaction along the singlet surface. The initial reactant and final product states have the same configuration weights for both the singlet and the triplet, explaining their equivalency in the table.

| Transition | λ(nm) | $A_{ik}$ (s$^{-1}$) | Laser Source |
|---|---|---|---|
| *4s4p* $^1$**P**$_1$← *4s$^2$* $^1$**S**$_0$ | 423 nm | 2.17 x 10$^8$ | Toptica TA-SHG 110 |
| *4s6f* $^1$**F**$_3$← *3d4s* $^1$**D**$_2$ | 411 nm | 2.74 x 10$^7$ | MSquared SOLSTIS ECD-X |
| *4p$^2$* $^3$**P**$_1$← *4s4p* $^3$**P**$_0$ | 429 nm | 6.11 x 10$^7$ | MSquared SOLSTIS ECD-X |
| *4p$^2$* $^3$**P**$_0$← *4s4p* $^3$**P**$_1$ | 431 nm | 1.82 x 10$^8$ | MSquared SOLSTIS ECD-X |
| *4s4d* $^3$**D**$_2$← *4s4p* $^3$**P**$_2$ | 446 nm | 1.88 x 10$^7$ | MSquared SOLSTIS ECD-X |
| *4s5p* $^1$**P**$_1$← *3d4s* $^1$**D**$_2$ | 672 nm | 1.30 x 10$^7$ | Homebuilt ECDL |

**Table S2: A description of the various laser transitions driven in this study to control the populations of Ca electronic states**.
For each transition used to control Ca electronic state populations, the initial and final electronic state, the transition wavelength, $\lambda$, the Einstein A coefficient ($A_{ik}$), and the source of the laser light used to drive each transition are presented. The Toptica TA-SHG 110 refers to an infrared diode laser system coupled to a tapered amplifier and second harmonic generation doubling cavity, while the MSquared SOLSTIS ECD-X is a titanium sapphire laser coupled to an external cavity frequency doubler. At any given time, at most two of the 411 nm, 429 nm, 431 nm, and 446 nm transitions needed to be driven, rendering the two available MSquared lasers in our system adequate for the experiment.